\documentclass[a4paper,11pt]{article}
\pdfoutput=1 
\usepackage{jcappub} 

\usepackage[T1]{fontenc}
\usepackage{mathrsfs}
\usepackage{amsmath}
\usepackage{upgreek}
\usepackage{multirow}
\usepackage{array}
\usepackage{booktabs}

\title{\boldmath Rational Orbits and Gravitational-Wave Signatures around a Horndeski Black Hole}

\author[a]{Wei Gao,}
\author[b]{Xiaoyan Zhu,}
\author[b,c,1]{Wenbin Lin,\note{Corresponding author.}}
\author[c]{and Siming Liu}

\affiliation[a]{School of Science, Hunan Institute Of Technology, \\ Hengyang 421001, People's Republic of China}
\affiliation[b]{School of Mathematics and Physics, University of South China, \\ Hengyang 421001, People's Republic of China}
\affiliation[c]{School of Physical Science and Technology, Southwest Jiaotong University, \\ Chengdu 610031, People's Republic of China}

\emailAdd{weigao2613@hnit.edu.cn}
\emailAdd{xyzhu0128@163.com}
\emailAdd{lwb@usc.edu.cn}
\emailAdd{liusm@swjtu.edu.cn}

\abstract{We study equatorial timelike geodesics and associated gravitational-wave (GW) emission for a static Horndeski black hole with secondary scalar hair, characterized by an effective pressure parameter $p_{\mathrm{eff}}$. Using the rational-orbit taxonomy $(z,w,v)$, we analyze the innermost stable circular orbit (ISCO), marginally bound orbit (MBO), and periodic bound orbits. We find that increasing $p_{\mathrm{eff}}$ systematically shifts orbital radii outward and modifies the rational parameter $q$. Within the test-particle and quadrupole approximations, we estimate GW signals from an extreme mass-ratio inspiral (EMRI) system at a luminosity distance of $D_L=8000\,\mathrm{pc}$. The characteristic strains reach $\sim10^{-18}$ in the mHz band, partially overlapping the nominal LISA sensitivity range. These results demonstrate that Horndeski corrections systematically alter strong-field orbital dynamics and GW spectra, offering a phenomenological probe of deviations from general relativity.}

\begin{document}
\maketitle
\flushbottom

\section{Introduction}\label{sec:intro}
Over the past decade, a series of landmark observations has opened a new era in the study of black holes and strong-field gravity. The first direct detection of gravitational waves from a binary black-hole merger by the Laser Interferometer Gravitational-Wave Observatory (LIGO)~\cite{LIGOScientific:2016aoc}, together with the horizon-scale observations of M87* and Sagittarius A* (Sgr A*) by the Event Horizon Telescope (EHT)~\cite{EventHorizonTelescope:2019dse,EventHorizonTelescope:2022wkp}, has provided unprecedented direct probes of black-hole systems. Gravitational-wave observations have also enabled an observational test of Hawking's black-hole area theorem~\cite{Isi:2020tac}, further demonstrating the potential of black holes as laboratories for testing fundamental predictions of gravitational theory. These advances have consequently motivated increasing interest in physical processes occurring in the immediate vicinity of black holes, where strong-field effects become significant and the dynamics of matter and test particles are determined directly by the underlying spacetime geometry~\cite{Berti:2015itd,Cardoso:2019rvt}.

Within the strong-field region surrounding a black hole, massive particles can exhibit a wide variety of orbital behaviors. Depending on the background spacetime, orbital parameters, and possible interactions with external fields, the motion may include bound and precessing trajectories, chaotic dynamics~\cite{Takahashi:2008zh,Kopacek:2014moa,Stuchlik:2015nlt,Li:2018wtz,Panis:2019urz,Hu:2021paa,Sun:2021oxg,Huang:2022qqu,Zhang:2023lrt,Xu:2024ble}, off-equatorial configurations~\cite{Kovar:2008ta,Kovar:2010ty}, and high-energy particle collisions in the near-horizon region~\cite{Banados:2009pr,Jacobson:2009zg,Wei:2010vca,Guo:2016vbt}. These orbital phenomena provide useful probes of the underlying spacetime geometry and of the strong-field interaction between gravity and matter. The orbital dynamics of test particles has also been investigated in a variety of non-Schwarzschild black-hole spacetimes, where additional charge or modified-gravity parameters can produce systematic modifications to the orbital period, periastron advance, and other orbital characteristics~\cite{Gao:2025lir,Gao:2025ufw,Gao:2025pdl}. Of particular interest are periodic orbits, which form an important organizing structure for bound orbital motion. Levin and Perez-Giz introduced a systematic taxonomy in which periodic timelike geodesics are classified through a correspondence with rational numbers~\cite{Levin:2008mq}. This framework provides a compact description of otherwise intricate zoom--whirl orbital configurations and has subsequently been extended and applied to a variety of black-hole spacetimes~\cite{Misra:2010pu,Grossman:2011ps}.

Among the various orbital structures in the strong-field regime, periodic orbits provide a particularly useful framework for characterizing bound black-hole dynamics. In Kerr spacetime, generic bound geodesic motion is characterized by the fundamental radial, polar, and azimuthal frequencies, $\Omega_r$, $\Omega_\theta$, and $\Omega_\varphi$~\cite{Schmidt:2002qk,Fujita:2009bp}. Periodic or resonant motion arises when the relevant fundamental frequencies become commensurate. For equatorial bound motion, Levin and Perez-Giz introduced a systematic taxonomy in which periodic orbits are associated with rational numbers and labeled by an integer triplet $(z,w,v)$, denoting the zoom number, whirl number, and vertex index, respectively~\cite{Levin:2008mq}. This rational-orbit framework provides a compact description of intricate zoom--whirl trajectories and establishes a direct connection between orbital morphology and the underlying spacetime geometry. It has subsequently been extended and applied to a variety of non-Kerr black holes, modified-gravity spacetimes, and alternative compact objects~\cite{Misra:2010pu,Liu:2018vea,Deng:2020yfm,Deng:2020hxw,Lin:2021noq,Gao:2020wjz,Lin:2022wda,Zhang:2022zox,Lin:2023rmo,
Shaymatov:2023jfa,Tu:2023xab,QiQi:2024dwc,Lim:2024mkb,Li:2026pfa}, demonstrating its usefulness for probing strong-field gravitational effects and constraining deviations from the Kerr geometry.

Rational orbits are not merely mathematical constructions, since their frequency commensurabilities encode information about the underlying spacetime geometry and provide an important organizing structure for strong-field orbital dynamics. More generally, frequency commensurabilities also play a significant role in the evolution of extreme-mass-ratio inspirals (EMRIs), in which a stellar-mass compact object gradually inspirals toward a massive black hole under gravitational radiation reaction. In generic Kerr EMRIs, transient orbital resonances occur when the radial and polar frequencies become commensurate. Near such resonances, the standard adiabatic approximation can break down, leading to finite changes in the orbital constants and corresponding dephasing of the gravitational-wave signal~\cite{Flanagan:2010cd,Berry:2016bit,Speri:2021psr,Tan:2026yjg}. These transient resonances are distinct from the equatorial rational orbits considered here, which are characterized by the commensurability between radial and azimuthal motion, but both phenomena illustrate the close connection between orbital-frequency structure and observable gravitational-wave signatures. EMRIs are expected to be important sources in the millihertz band and constitute major science targets for future space-based gravitational-wave observatories such as LISA~\cite{Robson:2018ifk,Caprini:2025mfr}, TianQin~\cite{TianQin:2015yph,Li:2024rnk}, and Taiji~\cite{Ruan:2018tsw}. Accurate modeling of periodic, resonant, and near-resonant orbital dynamics is therefore important for identifying strong-field signatures and testing gravitational physics with future space-based observations.

Motivated by these considerations, it is natural to ask how departures from the Schwarzs-child geometry modify the structure of rational orbits and whether such modifications can leave identifiable signatures in the associated gravitational radiation. The present study addresses this question for a static and spherically symmetric black-hole solution in the quartic square-root sector of Horndeski gravity, characterized by the metric function $f(r)=1-2m/r-8\pi p_{\rm eff}/r^2$. Although periodic and precessing particle motion has previously been investigated in other Horndeski black-hole backgrounds~\cite{Lin:2023rmo}, the rational-orbit structure and its associated gravitational-wave signatures for the present square-root Horndeski geometry have received comparatively less attention. The bound timelike geodesics are therefore analyzed systematically, including the characteristic circular orbits and the parameter region supporting bound motion, and the resulting periodic trajectories are classified within the rational-orbit framework. The corresponding gravitational-wave signals are then estimated in the extreme-mass-ratio and leading-order quadrupole approximations to examine how the Horndeski parameter modifies the orbital morphology and the associated spectral features. Comparison with the Schwarzschild limit provides a direct means of assessing the phenomenological imprint of the Horndeski correction on strong-field orbital dynamics and gravitational-wave emission.

The remainder of this paper is organized as follows. Section~\ref{Sec.II} introduces the Horndeski black-hole geometry and its horizon structure. Section~\ref{Sec.III} presents the equations governing timelike geodesic motion and the corresponding effective potential. Section~\ref{Sec.IV} discusses circular timelike orbits and the critical orbital parameters relevant to bound motion. Section~\ref{Sec.V} constructs and classifies rational orbits and examines their dependence on the Horndeski parameter. Section~\ref{Sec.VI} investigates the gravitational-wave signals associated with representative rational orbits in the extreme-mass-ratio regime. Finally, Sec.~\ref{Sec.VII} summarizes the main results and discusses their implications. Throughout this paper, geometrized units $G=c=1$ are adopted.

\section{Horndeski Black Hole Geometry and Horizon Structure}\label{Sec.II}
We consider a shift-symmetric subclass of Horndeski gravity, for which the action is invariant under a constant shift of the scalar field,
\begin{eqnarray}
&& \phi\rightarrow\phi+c,\label{c}
\end{eqnarray}
where $c$ is an arbitrary constant. Consequently, the Horndeski coupling functions have no explicit dependence on $\phi$ and depend only on the kinetic density
\begin{eqnarray}
&& X\equiv-\frac{1}{2}\nabla_\mu\phi\nabla^\mu\phi. \label{X}
\end{eqnarray}
The shift-symmetric Horndeski action can be written in the standard form~\cite{Horndeski:1974wa,Deffayet:2011gz}
\begin{eqnarray}
&& S = \int d^4x \sqrt{-g} \left(\mathcal{L}_2+\mathcal{L}_3+\mathcal{L}_4+\mathcal{L}_5\right),\label{S}
\end{eqnarray}

In the present work, we focus on the quartic square-root Horndeski model introduced in Ref.~\cite{Babichev:2017guv}, specified by
\begin{eqnarray}
&& G_2(X) = \eta X, \label{G2}\\
&& G_3(X) = 0, \label{G3}\\
&& G_4(X) = \zeta+\beta\sqrt{-X}, \label{G4}\\
&& G_5(X) = 0, \label{G5}
\end{eqnarray}
Here, $\eta$ and $\beta$ are coupling parameters, while $\zeta$ sets the coefficient of the Einstein--Hilbert term. The corresponding nonvanishing Horndeski Lagrangian densities are
\begin{eqnarray}
&& \mathcal{L}_2 = \eta X, \label{L2}\\
&& \mathcal{L}_3 = 0, \label{L3}\\
&& \mathcal{L}_4 = \left(\zeta+\beta\sqrt{-X}\right)R-\frac{\beta}{2\sqrt{-X}}\left[\left(\Box\phi\right)^2
-\left(\nabla_\mu\nabla_\nu\phi\right)\left(\nabla^\mu\nabla^\nu\phi\right)\right],\label{L4}\\
&& \mathcal{L}_5 = 0, \label{L5}
\end{eqnarray}
where
\begin{eqnarray}
&& \Box\phi\equiv\nabla_\mu\nabla^\mu\phi,\label{B}
\end{eqnarray}
and
\begin{eqnarray}
&& G_{4X}\equiv\frac{\partial G_4}{\partial X} = -\frac{\beta}{2\sqrt{-X}}.\label{G4X}
\end{eqnarray}

Following Ref.~\cite{Babichev:2017guv}, we take
\begin{eqnarray}
&& \zeta=\frac{M_{\rm Pl}^{2}}{16\pi}. \label{zeta}
\end{eqnarray} 
where the Planck-mass convention $M_{\rm Pl}^2=G^{-1}$ is used. Hence, in geometrized units $G=c=1$,
\begin{equation}
\zeta=\frac{1}{16\pi}.
\end{equation}

We seek a static and spherically symmetric black-hole solution with a static radial scalar field. The corresponding ansatz is
\begin{eqnarray}
&& ds^2=-h(r)dt^2+\frac{dr^2}{f(r)} +r^2\left(d\theta^2+\sin^2\theta\,d\varphi^2\right), \label{ansatz}\\ 
&& \phi=\phi(r). \label{scalaransatz}
\end{eqnarray} 

The shift symmetry gives rise to a conserved Noether current, $\nabla_\mu J^\mu=0$. For the static and spherically symmetric configuration considered here, only its radial component is nonvanishing. For the model defined above, it takes the form
\begin{eqnarray}
&& J^r=\frac{\beta\sqrt{2f}}{r^2}\,{\rm sgn}(\phi')-\eta f\phi'.\label{Jr}
\end{eqnarray}
Regularity of the current at the horizon selects the branch $J^r=0$~\cite{Babichev:2017guv}. For the nontrivial scalar-field configuration, this condition yields
\begin{eqnarray}
&& \phi'(r) =\pm\frac{\sqrt{2}\,\beta} {\eta r^2\sqrt{f(r)}}. \label{phiprime}
\end{eqnarray}

The corresponding kinetic density is therefore
\begin{eqnarray}
&& X(r)=-\frac{\beta^2}{\eta^2r^4}. \label{Xsolution}
\end{eqnarray}
Thus, $X<0$ in the static exterior region and the quantity $\sqrt{-X}$ appearing in the action is real. Consistency of Eq.~\eqref{Jr} on this nontrivial static branch further requires $\beta$ and $\eta$ to have the same sign~\cite{Babichev:2017guv}.

Substitution of Eq.~\eqref{phiprime} into the gravitational field equations determines the two metric functions. The $(tt)$ equation gives $f(r)$, while the $(rr)$ equation implies $h(r)=f(r)$ up to a constant rescaling of the time coordinate.
The resulting asymptotically flat solution is~\cite{Babichev:2017guv}
\begin{eqnarray}
&& ds^2 = -f\left(r\right)dt^2 + f^{-1}\left(r\right)dr^2 + r^2d\theta^2 + r^2\sin^2\theta d\varphi^2,\label{ds}
\end{eqnarray}
with
\begin{eqnarray}
&& f(r)=1-\frac{\mu}{r} -\frac{\beta^2}{2\zeta\eta r^2}. \label{foriginal}
\end{eqnarray}
The integration constant $\mu$ determines the ADM mass of the black hole and, in the geometrized units adopted here, is related to the black-hole mass $m$ by
\begin{eqnarray}
&& f(r)=1-\frac{2m}{r} -\frac{8\pi\beta^2}{\eta r^2}. \label{f}
\end{eqnarray}
For convenience, we introduce the effective Horndeski parameter
\begin{eqnarray}
&& p_{\rm eff}\equiv\frac{\beta^2}{\eta}, \label{peff}
\end{eqnarray}
Using $\zeta=1/(16\pi)$, Eq.~(\ref{foriginal}) becomes
\begin{eqnarray}
&& f(r)=1-\frac{2m}{r} -\frac{8\pi p_{\rm eff}}{r^2}. \label{fpeff}
\end{eqnarray}
The parameter $p_{\rm eff}$ is fixed by the couplings of the underlying Horndeski theory rather than by an independent integration constant. It therefore represents secondary scalar hair of the black hole. In the limit $p_{\rm eff}\rightarrow0$, the Schwarzschild geometry is recovered.

The horizon structure depends sensitively on the effective Horndeski parameter $p_{\rm eff}$. The roots of $f(r)=0$ are
\begin{eqnarray}
&& r_{\pm}=m\pm\sqrt{m^2+8\pi p_{\rm eff}}. \label{rpm}
\end{eqnarray}
For $p_{\rm eff}>0$, the solution possesses a single positive horizon radius, $r_+$, whereas the second root $r_-$ is negative and hence has no physical relevance. For $-m^2/(8\pi)<p_{\rm eff}<0$, both roots are positive, corresponding, at the level of the RN-like metric, to an outer horizon $r_+$ and an inner horizon $r_-$. At the critical value
\begin{eqnarray}
&& p_{\rm eff}=-\frac{m^2}{8\pi}, \label{peff}
\end{eqnarray}
the two roots coincide at $r=m$, giving an extremal geometry. For $p_{\rm eff}<-m^2/(8\pi)$, the metric function has no real zeros and the central singularity is naked. We note, however, that for the negative-$\eta$ branch associated with $p_{\rm eff}<0$, the static scalar-field configuration is well defined only in the region where $f(r)>0$, so the extension through the outer horizon requires additional care~\cite{Babichev:2017guv}.

\begin{figure}
  \centering
  \includegraphics[scale=0.65]{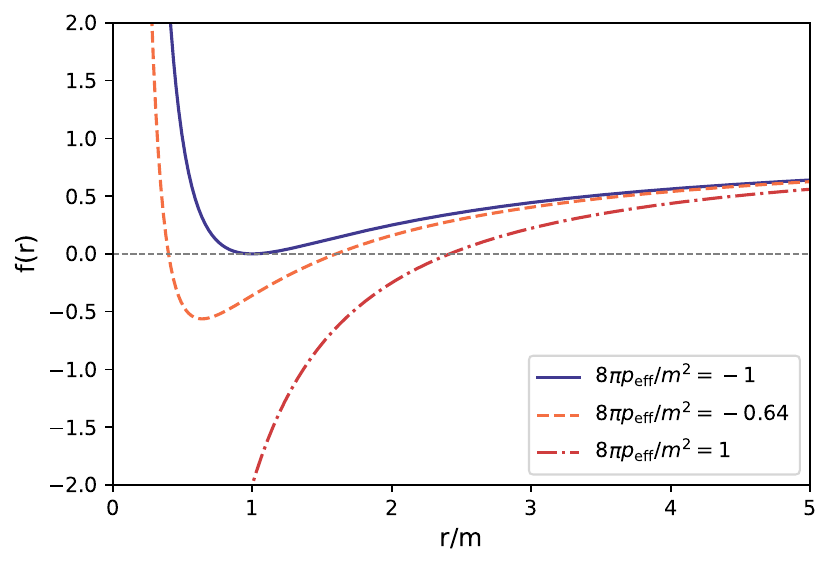}
  \caption{Metric function $f(r)$ for representative values of the effective Horndeski parameter $p_{\rm eff}$.}
  \label{figure1}
\end{figure}

Figure~\ref{figure1} illustrates these three characteristic horizon structures through the radial dependence of the metric function $f(r)$. The zeros of $f(r)$ determine the locations of the metric horizons.

The above classification determines the geometrically admissible parameter space considered in this work. We restrict our analysis to $p_{\rm eff}\geq -m^{2}/(8\pi)$, for which the spacetime possesses an event horizon, together with the condition $\beta\eta>0$ required by the nontrivial static scalar branch. Our primary interest is in the orbital dynamics of test particles and the corresponding gravitational-wave signatures, and the Horndeski geometry is therefore treated as a fixed background. We note that perturbative analyses of the square-root Horndeski model have reported stability issues near the event horizon~\cite{Minamitsuji:2022mlv}; a detailed stability analysis of the background itself is beyond the scope of the present study.

\section{Timelike Geodesics and Effective Potential}\label{Sec.III}
For the analysis of timelike geodesic motion, we first write the nonvanishing covariant components of the metric explicitly. In geometrized units, $G=c=1$, the line element introduced above gives
\begin{eqnarray}
&& g_{tt}=-f(r)=-\left(1-\frac{2m}{r}-\frac{8\pi p_{\rm eff}}{r^2}\right), \nonumber\\
&& g_{rr}=\frac{1}{f(r)}=\left(1-\frac{2m}{r}-\frac{8\pi p_{\rm eff}}{r^2}\right)^{-1}, \nonumber\\
&& g_{\theta\theta}=r^2, \nonumber\\
&& g_{\varphi\varphi}=r^2\sin^2\theta, \label{metric_components}
\end{eqnarray}
with all off-diagonal components vanishing.

We consider a massive test particle following a timelike geodesic, whose worldline is parametrized by the proper time $\tau$ as
\begin{eqnarray}
&& x^\mu(\tau)=\bigl(t(\tau),r(\tau),\theta(\tau),\varphi(\tau)\bigr),
\end{eqnarray}
where $\tau$ is the proper time along the particle worldline. The four-velocity is defined as
\begin{eqnarray}
&& u^\mu \equiv \frac{dx^\mu}{d\tau} = \dot{x}^\mu, \label{fourvelocity}
\end{eqnarray}
and satisfies the timelike normalization condition
\begin{eqnarray}
&& g_{\mu\nu}u^\mu u^\nu=-1. \label{normalization}
\end{eqnarray}

The geodesic motion can be described by the Lagrangian per unit rest mass,
\begin{eqnarray}
&& \mathcal{L}_p =\frac{1}{2}g_{\mu\nu}\dot{x}^{\mu}\dot{x}^{\nu}
= \frac{1}{2}\left[ g_{tt}\dot{t}^{\,2} +g_{rr}\dot{r}^{\,2} +g_{\theta\theta}\dot{\theta}^{\,2} +g_{\varphi\varphi}\dot{\varphi}^{\,2} \right]. \label{ML}
\end{eqnarray}
For proper-time parametrization, Eq.~\eqref{normalization} implies $\mathcal{L}_p=-1/2$.

Since the spacetime is stationary and spherically symmetric, the coordinates $t$ and $\varphi$ are cyclic. The associated conserved quantities are the specific energy $E$ and the azimuthal specific angular momentum $L$. Defining the conjugate momentum by
\begin{eqnarray}
&& p_\mu \equiv \frac{\partial\mathcal{L}}{\partial\dot{x}^{\mu}} =g_{\mu\nu}\dot{x}^{\nu}, \label{pmu}
\end{eqnarray}
we obtain
\begin{eqnarray}
&& p_t=g_{tt}\dot{t}=-E,\qquad p_\varphi=g_{\varphi\varphi}\dot{\varphi}=L. \label{conserved}
\end{eqnarray}
Therefore,
\begin{eqnarray}
&& \dot{t}=\frac{E}{f(r)},\qquad  \dot{\varphi}=\frac{L}{r^2\sin^2\theta}.\label{tdotphidot}
\end{eqnarray}
while the remaining momenta are
\begin{eqnarray}
&& p_r=g_{rr}\dot{r}=\frac{\dot{r}}{f(r)},\qquad p_\theta=g_{\theta\theta}\dot{\theta} =r^2\dot{\theta}.\label{rt}
\end{eqnarray}

Performing a Legendre transformation, the Hamiltonian per unit rest mass is
\begin{eqnarray}
&& \mathcal{H}=p_\mu\dot{x}^{\mu}-\mathcal{L}=\frac{1}{2}g^{\mu\nu}p_\mu p_\nu,\label{H}
\end{eqnarray}
where $g^{\mu\nu}$ is the inverse metric satisfying $g^{\mu\alpha}g_{\alpha\nu}=\delta^\mu{}_\nu$. For the present spacetime, the Hamiltonian can be written explicitly as
\begin{eqnarray}
&& \mathcal{H} =\frac{1}{2}\left[-\frac{E^2}{f(r)}+f(r)p_r^2+\frac{p\theta^2}{r^2}+\frac{L^2}{r^2\sin^2\theta}\right]. \label{Hexplicit}
\end{eqnarray}
For a massive particle parametrized by proper time, the normalization condition is equivalent to the Hamiltonian constraint
\begin{eqnarray}
&& \mathcal{H}=-\frac{1}{2}. \label{Hconstraint}
\end{eqnarray}

Owing to spherical symmetry, the orbital plane can always be chosen to coincide with the equatorial plane without loss of generality. Hence, we set $\theta=\frac{\pi}{2}$, $\dot{\theta}=0$, $p_\theta=0$. On the equatorial plane,
\begin{eqnarray}
&& \dot{\varphi}=\frac{L}{r^2}.
\end{eqnarray}

Using the normalization condition together with the conserved quantities $E$ and $L$, one obtains
\begin{eqnarray}
&& \dot{r}^{\,2} +\frac{1}{g_{rr}} \left(1+\frac{L^2}{g_{\varphi\varphi}}\right)=-\frac{E^2}{g_{tt}g_{rr}}. \label{radial_general}
\end{eqnarray}
For the metric considered here,
\begin{eqnarray}
&& g_{tt}g_{rr}=-1, \qquad g_{\varphi\varphi}=r^2
\end{eqnarray}
on the equatorial plane. Equation~(\ref{radial_general}) therefore takes the standard form
\begin{eqnarray}
&& \dot{r}^{\,2}+V_{\rm eff}(r)=E^2, \label{radial_equation}
\end{eqnarray}
where the effective potential for timelike geodesic motion is
\begin{eqnarray}
&& V_{\rm eff}(r) = f(r)\left(1+\frac{L^2}{r^2}\right) = \left(1-\frac{2m}{r}-\frac{8\pi p_{\rm eff}}{r^2}\right)\left(1+\frac{L^2}{r^2}\right).\label{Veff}
\end{eqnarray}

To investigate how the Horndeski correction affects the orbital dynamics, we analyze the effective potential for different values of $p_{\rm eff}$. The extrema and stability properties of $V_{\rm eff}(r)$ determine the characteristic circular orbits of the spacetime. The corresponding critical orbital parameters, including the innermost stable circular orbit and the marginally bound orbit, will be investigated in the following section.

\section{Circular Orbits and Critical Orbital Parameters}\label{Sec.IV}
For circular equatorial geodesics, the radial velocity vanishes, $\dot r=0$, and the orbital radius corresponds to an extremum of the effective potential $V_{\rm eff}(r)=f(r)\bigl(1+L^2/r^2\bigr)$ given in Eq.~\eqref{Veff}. Among the circular geodesics, two critical configurations are particularly relevant to the bound orbital dynamics considered in this work.

\paragraph{Innermost stable circular orbit (ISCO).}
The ISCO is the innermost circular geodesic that remains stable against radial perturbations. It is determined by
\begin{eqnarray}
&& V_{\rm eff}(r)=E^2,\qquad \frac{dV_{\rm eff}}{dr}=0,\qquad \frac{d^2V_{\rm eff}}{dr^2}=0 . \label{ISCOcond}
\end{eqnarray}
Solving Eq.~\eqref{ISCOcond} yields the critical radius $r_{\rm ISCO}$, together with the corresponding specific angular momentum $L_{\rm ISCO}$ and specific energy $E_{\rm ISCO}$.

\paragraph{Marginally bound orbit (MBO).}
The MBO is an unstable circular orbit with vanishing binding energy, corresponding to $E=1$. It therefore satisfies
\begin{eqnarray}
&& V_{\rm eff}(r)=1,\qquad \frac{dV_{\rm eff}}{dr}=0 ,\label{MBOcond}
\end{eqnarray}
without imposing the marginal-stability condition. The MBO represents the limiting circular configuration separating bound motion from marginally unbound motion.

Table~\ref{table1} collects the resulting critical orbital parameters for three representative values of the Horndeski parameter. Both $r_{\rm ISCO}$ and $r_{\rm MBO}$ increase monotonically as $8\pi p_{\rm eff}/m^2$ increases from $-0.5$ to $+0.5$, recovering the Schwarzschild values, $r_{\rm ISCO}=6m$ and $r_{\rm MBO}=4m$, at $p_{\rm eff}=0$. The corresponding critical angular momenta exhibit the same monotonic trend. These shifts reflect the modification of the strong-field effective potential induced by the Horndeski $1/r^2$ term.

\begin{table}[htbp]
    \centering
    \caption{Critical orbital parameters for timelike particles. All quantities are given in geometric units ($G=c=1$) and normalized by the black-hole mass $m$ where appropriate.}
    \label{table1}
    \setlength{\tabcolsep}{10pt} 
    \begin{tabular}{ccccc}
        \toprule \hline
        $8\pi p_{\rm eff}/m^2$ & $r_{\rm ISCO}/m$ & $r_{\rm MBO}/m$ & $L_{\rm ISCO}/m$ & $L_{\rm MBO}/m$ \\ \hline
        \midrule 
        $-0.5$ & 5.1665 & 3.4397 & 3.1962 & 3.7207 \\ \hline
        $0$    & 6.0000 & 4.0000 & 3.4641 & 4.0000 \\ \hline
        $0.5$  & 6.6945 & 4.4609 & 3.6859 & 4.2301 \\ \hline
        \bottomrule
    \end{tabular}
\end{table}

To parametrize bound precessing orbits, we introduce
\begin{eqnarray}
&& L_i=(1-\varepsilon)L_{\rm ISCO}+\varepsilon L_{\rm MBO},\qquad \varepsilon\in(0,1) , \label{epsi}
\end{eqnarray}
where $L_{\rm ISCO}$ and $L_{\rm MBO}$ are taken from Table~\ref{table1}. For a fixed $L_i$ in the interval $L_{\rm ISCO}<L_i<L_{\rm MBO}$, the effective potential generally possesses two extrema outside the event horizon: a local maximum at $r=r_u$, corresponding to an unstable circular orbit, and a local minimum at $r=r_s$, corresponding to a stable circular orbit. The associated critical energies are defined by
\begin{eqnarray}
&& E_{\max}=\sqrt{V_{\rm eff}(r_u)},\qquad E_{\min}=\sqrt{V_{\rm eff}(r_s)}. \label{Eminmax}
\end{eqnarray}
Noncircular bound motion is possible when
\begin{eqnarray}
&& E_{\min}<E_i<E_{\max}. \label{Einterval}
\end{eqnarray}

For such an energy, the radial turning points are determined by
\begin{eqnarray}
&& E_i^2-V_{\rm eff}(r)=0. \label{turningcondition}
\end{eqnarray}
In general, Eq.~\eqref{turningcondition} possesses three positive roots in the relevant strong-field regime. The innermost root belongs to the plunging branch, whereas the two outer roots correspond to the periastron $r_p$ and apastron $r_a$ of the bound orbit. The particle therefore oscillates radially between $r_p$ and $r_a$.

For the representative choice $\varepsilon=0.5$, Table~\ref{table2} lists the locations $r_u$ and $r_s$ of the two extrema of the effective potential, together with the corresponding critical energies $E_{\max}$ and $E_{\min}$ and the angular momentum $L_i$.
\begin{table}[htbp]
    \centering
    \caption{Characteristic extrema of the effective potential for equatorial timelike motion at $\varepsilon=0.5$. The radii $r_u$ and $r_s$ denote the unstable and stable circular orbits, respectively.}
    \label{table2}
    \setlength{\tabcolsep}{10pt} 
    \begin{tabular}{cccccc}
        \toprule \hline
        $8\pi p_{\mathrm{eff}}/m^2$ & $r_u$ & $r_p$ & $E_{\mathrm{max}}$ & $E_{\mathrm{min}}$ & $L_{\mathrm{i}}$ \\ \hline
        \midrule 
        $-0.5$ & 3.7566 & 8.3217 & 0.9640 & 0.9483 & 3.4585\\ \hline
        $0$    & 4.3729 & 9.5553 & 0.9684 & 0.9546 & 3.7321 \\ \hline
        $0.5$  & 4.8821 & 10.5869 & 0.9714 & 0.9588 & 3.9580 \\ \hline
        \bottomrule
    \end{tabular}
\end{table}

\begin{figure}
  \centering
  \includegraphics[scale=0.65]{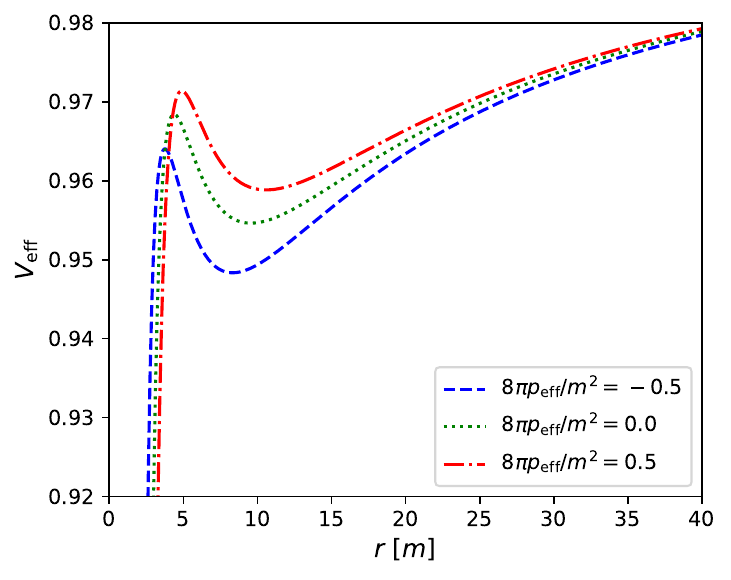}
  \caption{Effective potential $V_{\rm eff}$ as a function of the radial coordinate $r$ for $\varepsilon=0.5$ and different values of the Horndeski parameter. The local maximum and minimum correspond to the unstable and stable circular orbits, respectively.}
  \label{figure2}
\end{figure}

Figure~\ref{figure2} shows $V_{\rm eff}(r)$ for $\varepsilon=0.5$ and the three representative values of $p_{\rm eff}$. In each case, the exterior effective potential exhibits a local maximum at the unstable circular orbit $r_u$ and a local minimum at the stable circular orbit $r_s$. As $8\pi p_{\rm eff}/m^2$ increases, both characteristic radii shift outward. The corresponding critical energies $E_{\min}$ and $E_{\max}$ delimit the allowed energy interval for noncircular bound motion at the specified value of $L_i$.

\begin{figure}
  \centering
  \includegraphics[width=15cm]{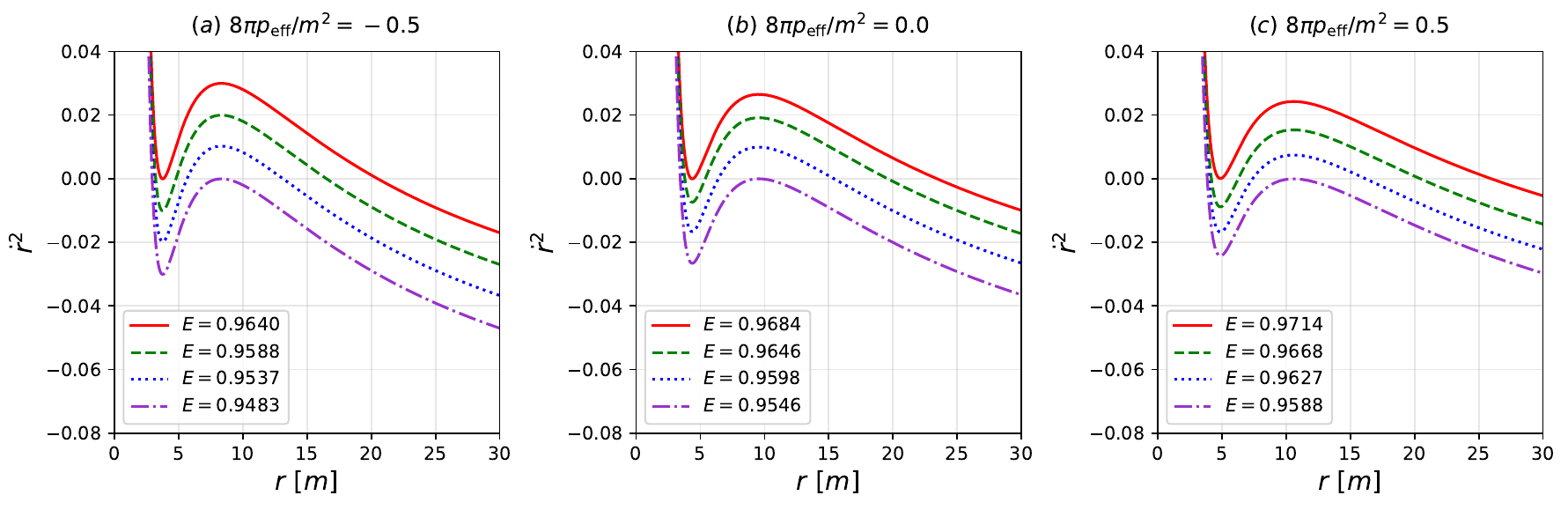}
  \caption{$\dot{r}^2=E^2-V_{\rm eff}(r)$ as a function of $r$ for $\varepsilon=0.5$. The solid and dash-dotted curves correspond to $E_{\rm max}$ and $E_{\rm min}$, respectively.}
  \label{figure3}
\end{figure}

The radial dynamics can be further illustrated through
\begin{eqnarray}
&& \dot r^2=E^2-V_{\rm eff}(r). \label{radialfunction}
\end{eqnarray}
Figure~\ref{figure3} shows $\dot r^2$ evaluated at the two limiting energies $E_{\max}$ and $E_{\min}$ for $\varepsilon=0.5$. At $E=E_{\max}$, the radial function develops a double root at $r=r_u$, corresponding to the unstable circular orbit and the separatrix between bound and plunging motion. At $E=E_{\min}$, a double root occurs at $r=r_s$, corresponding to the stable circular orbit.

For an intermediate energy $E_{\min}<E_i<E_{\max}$, these degenerate roots split. The radial equation then possesses an inner root associated with the plunging branch and two outer turning points $r_p$ and $r_a$ defining the bound orbit. As $E_i$ approaches $E_{\max}$ from below, the periastron $r_p$ approaches the unstable circular radius $r_u$, and the particle spends an increasingly long time near $r_u$ before zooming back toward the apastron. This behavior provides the dynamical origin of the zoom-whirl structure discussed below.

\begin{figure}
  \centering
  \includegraphics[width=15cm]{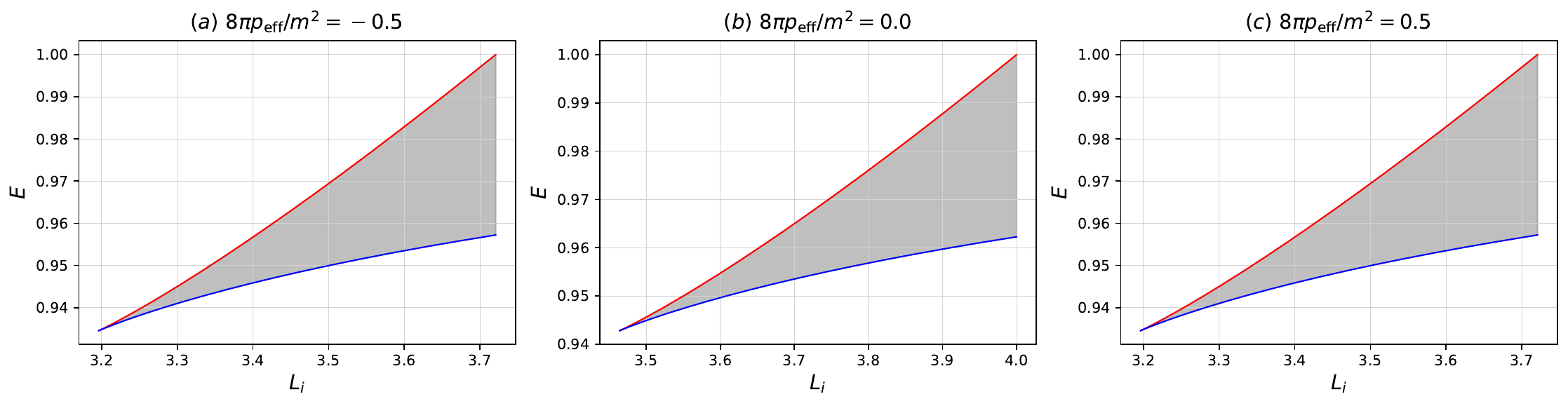}
  \caption{Evolution of the maximum specific energy $E_{\mathrm{max}}$ and the minimum specific energy $E_{\mathrm{min}}$ corresponding to stable orbits with respect to the orbital angular momentum $L_i$. The shaded region between the two curves represents the parameter space where stable orbits exist.}
  \label{figure4}
\end{figure}

Figure~\ref{figure4} shows the critical energies associated with the local minimum, $E_{\min}(L)$, and local maximum, $E_{\max}(L)$, of the effective potential as functions of the specific angular momentum $L$. For each value of $p_{\rm eff}$, the region $E_{\min}(L)<E<E_{\max}(L)$ and $L_{\rm ISCO}<L<L_{\rm MBO}$ defines the parameter space in which noncircular bound motion exists.

The two critical-energy branches merge at $L=L_{\rm ISCO}$, where the stable and unstable circular orbits coalesce and the marginal-stability condition $\partial_r^2V_{\rm eff}=0$ is satisfied. At the opposite end, $L=L_{\rm MBO}$, the upper branch reaches $E_{\max}=1$, corresponding to the marginally bound circular orbit. For fixed $L$, increasing $p_{\rm eff}$ lowers the values of the effective potential at its extrema and hence lowers both $E_{\min}$ and $E_{\max}$. This fixed-$L$ behavior should be distinguished from the values listed in Table~\ref{table2}, where $L_i$ itself varies with $p_{\rm eff}$ through Eq.~\eqref{epsi}.

Within the allowed region, energies close to $E_{\min}$ correspond to small radial oscillations around the stable circular orbit $r_s$. As the energy increases toward $E_{\max}$, the inner turning point of the bound orbit moves toward the unstable circular orbit $r_u$, giving rise to increasingly pronounced zoom-whirl behavior. The region near the upper boundary $E_{\max}(L)$ is therefore particularly relevant for the rational orbits investigated in the following section.

\section{Rational Orbits}\label{Sec.V}
Having established the parameter region for bound precessing motion, we now identify the closed periodic orbits contained within this region. We adopt the classification scheme introduced by Levin and Perez-Giz~\cite{Levin:2008mq}, hereafter referred to as the LPG taxonomy, in which periodic bound timelike geodesics are characterized by an integer triplet $(z,w,v)$ according to their zoom--whirl--vertex structure:
\begin{itemize}
  \item \textbf{Zoom number $z$}: counts the number of distinct radial leaves traced during one complete periodic orbit. Equivalently, $z$ gives the number of radial periods required for the orbit to close. During these $z$ radial periods, the particle reaches $z$ distinct apastron directions.

  \item \textbf{Whirl number $w$}: counts the additional quasi-circular azimuthal revolutions executed near periastron during each radial period. Thus, $w$ characterizes the strength of the whirl behavior in the strong-field region.

  \item \textbf{Vertex index $v$}: specifies which apastron vertex is reached after one radial period. For $z>1$, it satisfies $1\leq v\leq z-1$ and $\gcd(z,v)=1$, while for a single-leaf orbit, $z=1$, one has $v=0$. In the numerical examples below, we choose $v=z-1$ for $z>1$.
\end{itemize}

The LPG rational parameter is defined as
\begin{eqnarray}
&& q = w + \frac{v}{z},\label{q1}
\end{eqnarray}
Let $\Delta\varphi_r$ denote the accumulated azimuthal angle during one complete radial period, i.e., from one apastron through periastron to the next apastron. The rational parameter can then be written as
\begin{eqnarray}
&& q=\frac{\Delta\varphi_r}{2\pi}-1. \label{q2}
\end{eqnarray}
Combining Eqs.~\eqref{q1} and~\eqref{q2} gives
\begin{eqnarray}
&& \Delta\varphi_r =2\pi(1+q) = 2\pi\left(1+w+\frac{v}{z}\right). \label{deltaphir}
\end{eqnarray}
Since a closed $(z,w,v)$ orbit contains $z$ radial periods, the total azimuthal angle accumulated over one complete periodic orbit is
\begin{eqnarray}
&& \Delta\varphi_{\rm tot} = z\Delta\varphi_r =2\pi\left[z(1+w)+v\right]. \label{deltaphitot}
\end{eqnarray}
A rational value of $q$ corresponds to a closed periodic orbit, whereas an incommensurate radial--azimuthal frequency ratio leads to a nonclosing precessing trajectory.

On the equatorial plane, $\theta=\pi/2$, the azimuthal advance over one radial period can be obtained directly from the geodesic equations. Denoting the periastron and apastron by $r_p$ and $r_a$, respectively, we have
\begin{eqnarray}
&& \Delta\varphi_r = 2\int_{r_p}^{r_a} \frac{\dot{\varphi}}{\dot r}dr \nonumber\\
&& \hskip 0.7cm = 2\int_{r_p}^{r_a} \frac{L_i} {r^2\sqrt{ E_i^2- \left(1-\frac{2m}{r}-\frac{8\pi p_{\rm eff}}{r^2}\right)
\left(1+\frac{L_i^2}{r^2}\right) }}dr . \label{Dvarphi}
\end{eqnarray}
The radial turning points $r_p$ and $r_a$ are determined by
\begin{eqnarray}
&& E_i^2-V_{\rm eff}(r)=0. \label{turningpointq}
\end{eqnarray}

Equations~\eqref{q2} and~\eqref{Dvarphi} give the explicit map between the triplet $(z,w,v)$ and the conserved pair $(E_i,L_i)$ for fixed $p_{\mathrm{eff}}$. The proper time from periastron to the next apastron is
\begin{eqnarray}
&& \Delta\tau = \int_{r_p}^{r_a}\frac{1}{\sqrt{E_i^2-\left(1-\frac{2m}{r}-\frac{8\pi p_{\rm eff}}{r^2}\right)\left(1+\frac{L_i^2}{r^2}\right)}}dr,\label{tau}
\end{eqnarray}
Since one complete radial period consists of motion from $r_p$ to $r_a$ and back to $r_p$, a complete rational orbit comprising $z$ radial periods has the proper-time period $T_q = 2z\Delta\tau$.

\begin{figure}
  \centering
  \includegraphics[scale=0.65]{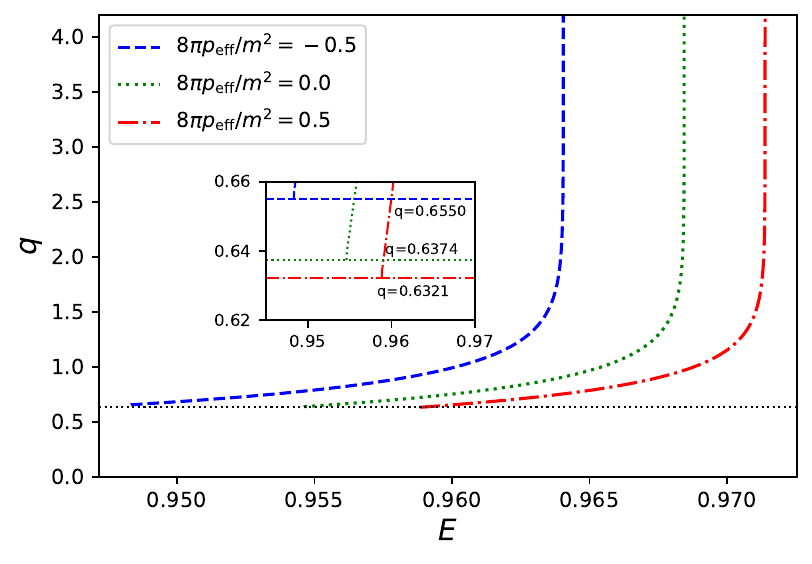}
  \caption{Rational parameter $q$ as a function of the specific energy $E_i$ at fixed $L_i$ with $\varepsilon=0.5$.}
  \label{figure5}
\end{figure}
Figure~\ref{figure5} shows the rational parameter $q=\Delta\varphi_r/(2\pi)-1$ as a function of the specific energy $E_i$ at fixed $L_i$ with $\varepsilon=0.5$. For all three representative values of $p_{\rm eff}$, $q$ increases monotonically with $E_i$. The minimum values are $q_{\min}\simeq0.6550,\,0.6374,\,0.6321$ for $8\pi p_{\rm eff}/m^{2}=-0.5,\,0,\,+0.5$, respectively. For the periodic family considered below with $w=1$, the single-leaf orbit $(z,w,v)=(1,1,0)$ has $q=1$, whereas for $z>1$ and $v=z-1$, one has $q=2-1/z$.

\begin{figure}
  \centering
  \includegraphics[width=14cm]{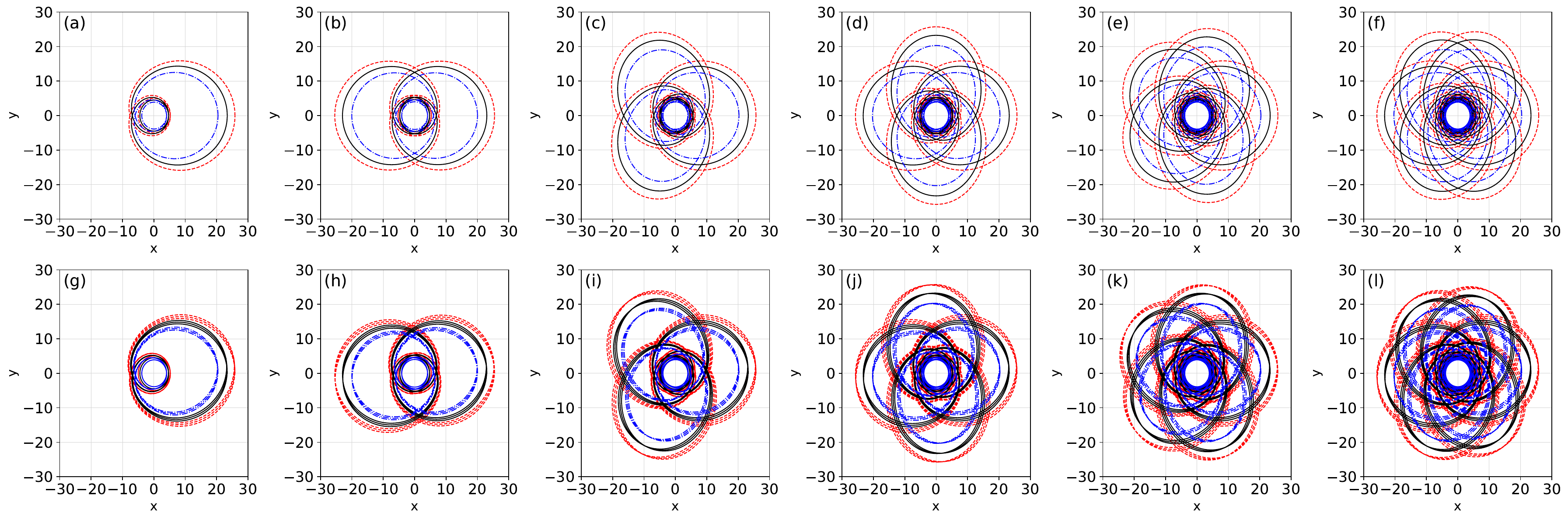}
  \caption{Rational (top) and irrational (bottom) bound equatorial timelike orbits, with the orbital lobe index $z$ increasing left to right from $1$ to $6$. The azimuthal orientation index $v$ satisfies $v=z-1$ for $z>1$ and $v=1$ for $z=1$. The red, black, and blue curves correspond to $8\pi p_{\mathrm{eff}}/m^2= 0.5$, $0.0$, $-0.5$ respectively.}
  \label{figure6}
\end{figure}

With $w=1$, we integrate bound timelike geodesics for $z\in\{1,\ldots,6\}$, taking $v=0$ for $z=1$ and $v=z-1$ for $z>1$. Each rational orbit is initialized at the rightmost apastron and evolved over its complete closed period, which contains $z$ radial periods. The resulting trajectories exhibit exactly $z$ distinct radial leaves, as shown in the top row of Fig.~\ref{figure6}. For comparison, the bottom row shows nearby nonperiodic trajectories obtained by slightly shifting the orbital parameters away from exact rational commensurability.

\section{Gravitational Waves from Rational Orbits}\label{Sec.VI}
The gravitational-wave strain associated with the bound equatorial geodesics derived above is evaluated within the test-particle and leading-order quadrupole approximations. The resulting waveforms are used to characterize the qualitative strain modulation and spectral features associated with rational orbital motion, rather than to provide fully self-consistent strong-field waveforms.

Within the Newtonian quadrupole point-particle formalism of Ref.~\cite{Babak:2006uv}, the plus and cross polarizations for an equatorial orbit read
\begin{eqnarray}
&& h_{+}=-\frac{2\eta}{D_{L}}\frac{G m}{c^{2} r}\left(1+\cos^{2}\iota\right)\cos(2\varphi+2\zeta),\label{h+}\\
&& h_{\times}=-\frac{4\eta}{D_{L}}\frac{G m}{c^{2} r}\cos\iota\,\sin(2\varphi+2\zeta).\label{ht}
\end{eqnarray}
Here, $\iota$ is the inclination angle between the orbital angular momentum and the line of sight, $\zeta$ denotes the orientation of the periastron, $D_L$ is the luminosity distance, and $\eta = m m'/(m+m')^{2}$ is the symmetric mass ratio, with $m'$ denoting the mass of the orbiting compact object.

Equations~\eqref{h+}--\eqref{ht} provide only a leading-order quadrupole estimate and should not be interpreted as self-consistent strong-field waveforms. In general relativity, accurate EMRI waveforms in the strong-field regime require black-hole perturbation theory together with gravitational self-force corrections~\cite{Hughes:1999bq,Drasco:2005kz,Pound:2012nt}. For the Horndeski black-hole background considered here, a fully consistent treatment would additionally require the corresponding modified-gravity perturbation equations, including possible couplings between tensor and scalar degrees of freedom. Accordingly, the present approximation is used only to characterize the qualitative waveform modulation and spectral signatures associated with rational orbital motion, while a complete strong-field waveform calculation lies beyond the scope of the present study.

For the numerical calculations, the central Horndeski black hole is assigned a mass $m=4\times10^{6}\,M_{\odot}$, with $p_{\mathrm{eff}}=0$ corresponding to the Schwarzschild limit, while the orbiting compact object has a mass $m'=100\,M_{\odot}$. The corresponding symmetric mass ratio is $\eta=mm'/(m+m')^{2}\simeq2.50\times10^{-5}$. Restoring $G$ and $c$ and adopting a source distance $D_L=8000\,\mathrm{pc}$, the characteristic strain scale is $\eta Gm/(D_Lc^{2})\simeq5.99\times10^{-16}$, while the characteristic gravitational time scale is $Gm/c^{3}\simeq19.71\,\mathrm{s}$. All orbital configurations considered below are evaluated at $\varepsilon=0.5$.

\begin{figure}
    \centering
    \includegraphics[width=15cm]{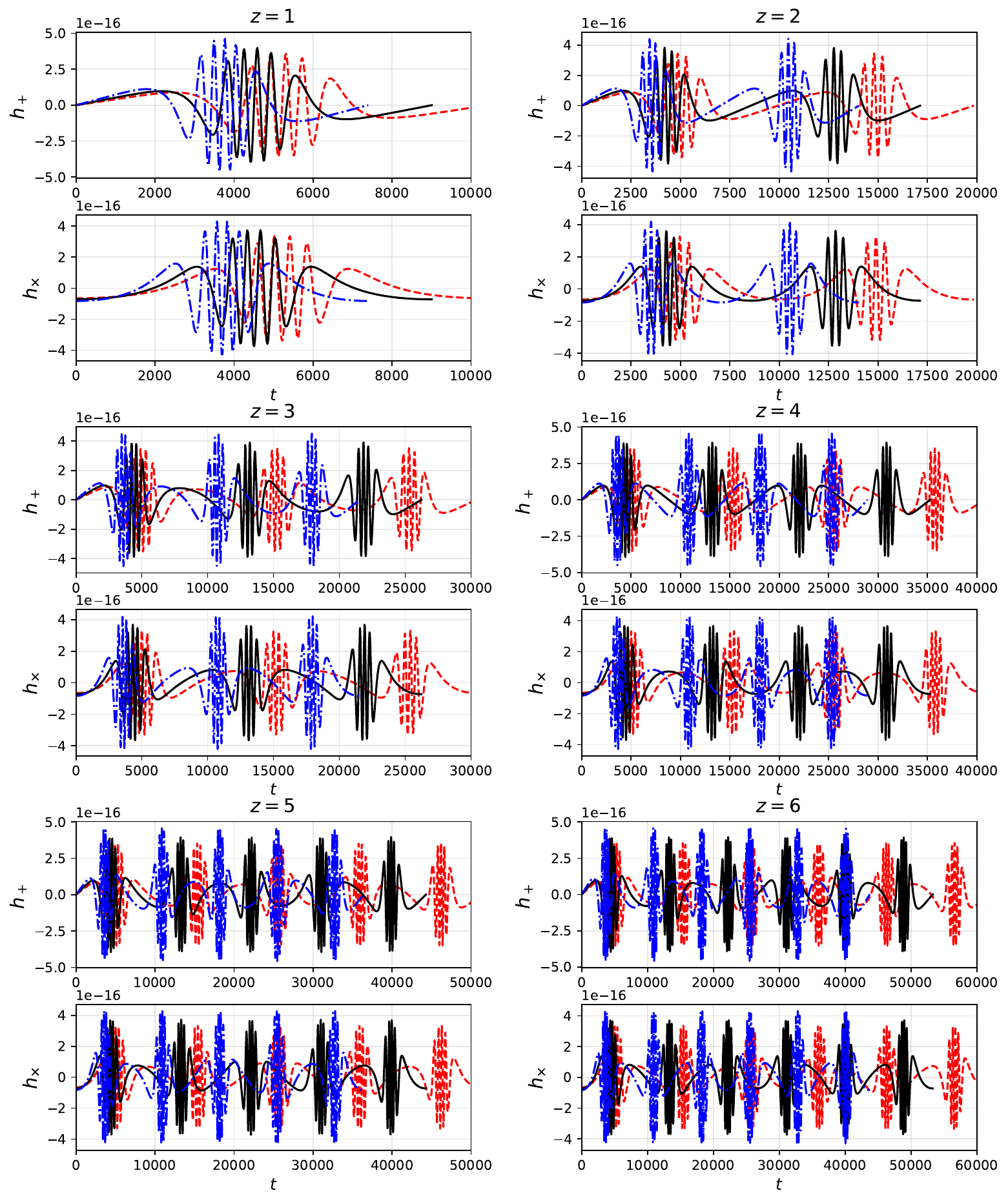}
    \caption{Time-domain gravitational wave signals for rational orbits. The parameter $z$ increases from $1$ to $6$ (arranged top-to-bottom, left-to-right). In each panel, the red, black, and blue curves represent $8\pi p_{\mathrm{eff}}/m^2 = 0.5$, $0.0$, $-0.5$, respectively.}
    \label{figure7}
\end{figure}

Figure~\ref{figure7} shows the time-domain polarizations $h_{+}(t)$ and $h_{\times}(t)$ for rational orbits with $z=1,\ldots,6$. Since each rational orbit is periodic with period $T_q$, the corresponding gravitational-wave signal inherits the periodic radial and azimuthal structure of the orbital motion.
\begin{figure}
  \centering
  \includegraphics[width=15cm]{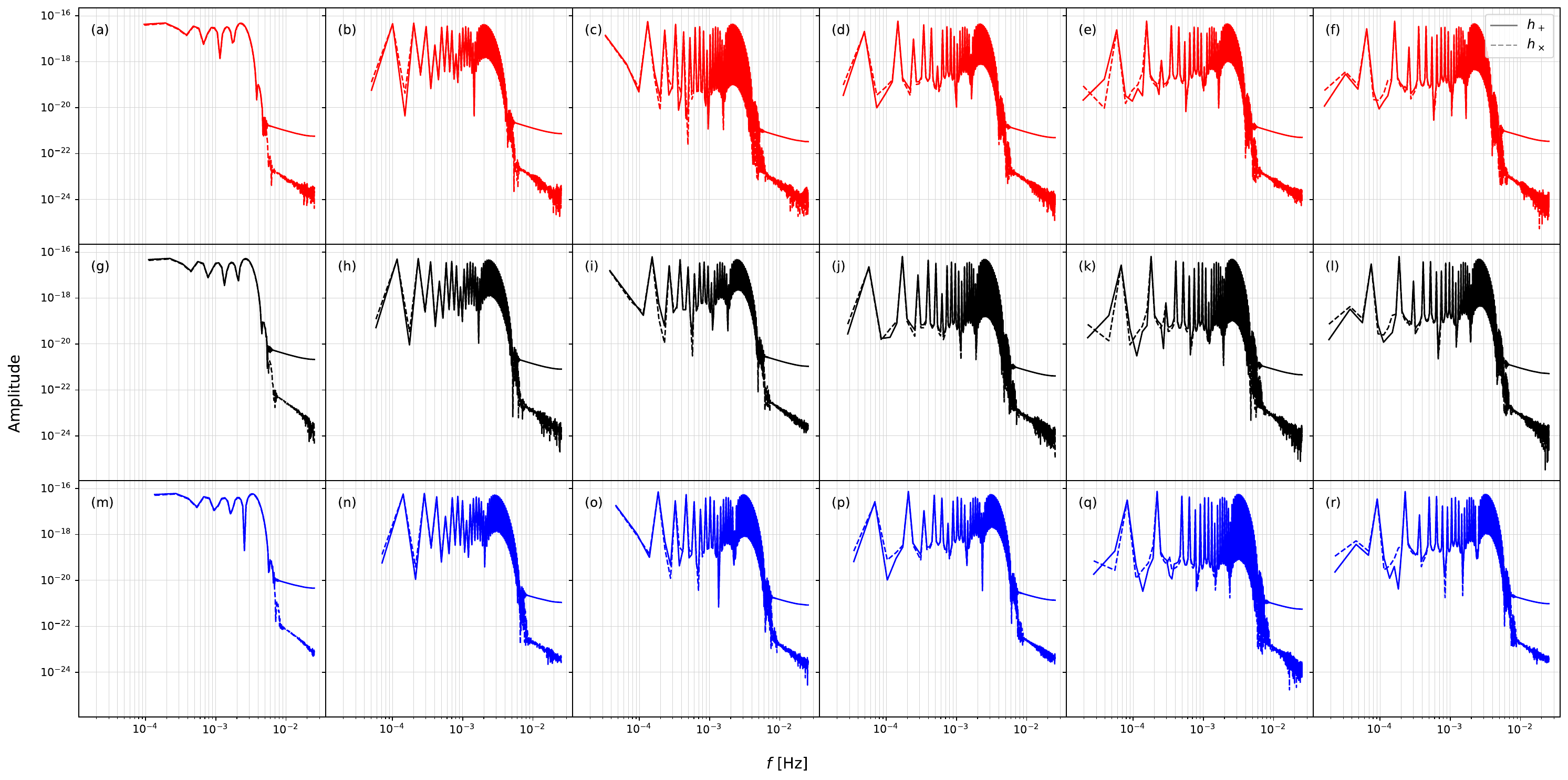}
  \caption{Frequency-domain gravitational-wave amplitudes for rational orbits with $z=1,\ldots,6$, arranged from left to right. The red, black, and blue curves correspond to $8\pi p_{\rm eff}/m^2=0.5$, $0$, and $-0.5$, respectively. The dominant spectral features are concentrated mainly below $0.01,\mathrm{Hz}$.}
  \label{figure8}
\end{figure}

The frequency-domain amplitudes $\hat h_+(f)$ and $\hat h_\times(f)$ are obtained by Fourier transforming the corresponding time-domain signals. Following Ref.~\cite{Finn:2000sy}, the characteristic strain is defined as
\begin{eqnarray}
&& h_c\left(f\right) = 2f\sqrt{\left |\hat{h}_+\left(f\right)\right|^2 +\left|\hat{h}_{\times}\left(f\right)\right|^2}.\label{hc}
\end{eqnarray}

\begin{figure}[htpb!]
  \centering
  \includegraphics[width=0.50\textwidth]{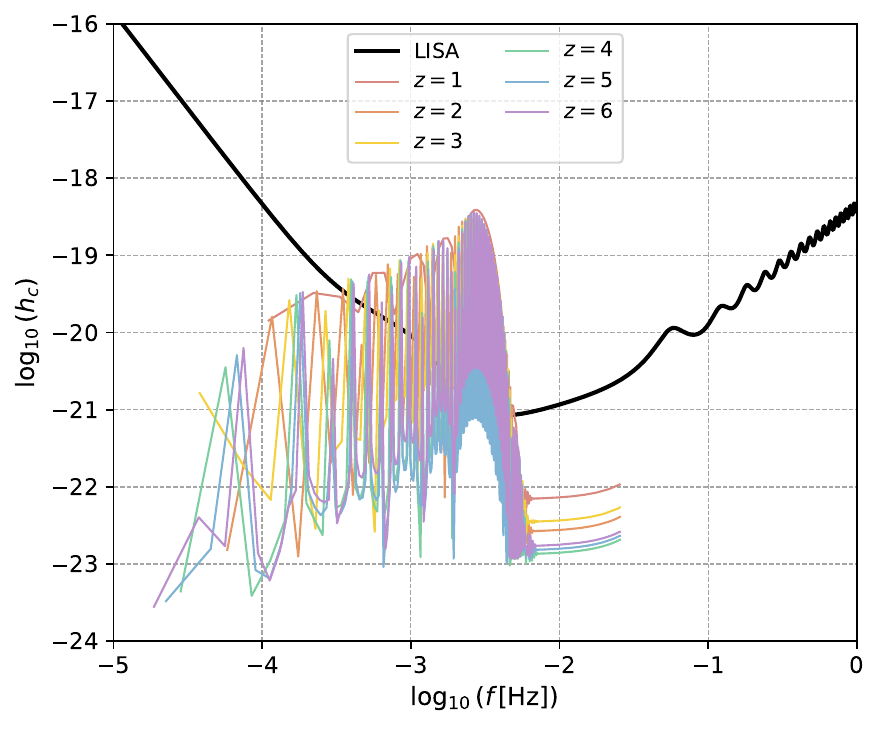}
  \caption{Characteristic strain spectra for various rational-orbit configurations (colored), overlaid with the sky-averaged LISA sensitivity curve (solid black)~\cite{Robson:2018ifk}. The signals overlap with or exceed the LISA sensitivity curve over part of the $\mathrm{mHz}$ band, indicating potential observability, although a quantitative detectability assessment requires an explicit SNR calculation.}
  \label{figure9}
\end{figure}

Figure~\ref{figure8} shows the frequency-domain amplitudes of the gravitational-wave signals, while Fig.~\ref{figure9} presents the corresponding characteristic strain $h_c(f)$ together with the sky-averaged LISA sensitivity curve for a 1-yr observation~\cite{Robson:2018ifk}. The signals are concentrated mainly in the mHz band, with the characteristic frequency scale set by the orbital period $T_q$ and its harmonics. For the source parameters adopted here, the characteristic strain overlaps the LISA sensitivity curve over part of this frequency range. This comparison provides only an indication of the relevant strain and frequency scales; a quantitative assessment of detectability would require a full signal-to-noise-ratio calculation including the detector response, observation time, confusion noise, and an accurate strong-field waveform model. In the present examples, increasing $8\pi p_{\rm eff}/m^2$ shifts the spectral features toward slightly lower frequencies, consistent with the outward displacement of the characteristic orbital radii and the corresponding increase in the orbital period.

Figure~\ref{figure9} compares the characteristic strain spectra with the sky-averaged LISA sensitivity curve. The dominant signal power lies in the mHz band, where part of the spectra overlaps with or exceeds the nominal LISA sensitivity level. This suggests potential observability, although a quantitative detectability assessment requires a full signal-to-noise-ratio calculation including the detector response, observation time, confusion noise, and waveform accuracy.

\section{Summary}\label{Sec.VII}

The bound equatorial timelike geodesics and their associated gravitational-wave signatures have been investigated for a static Horndeski black hole with secondary scalar hair, described by the metric function $f(r)=1-2m/r-8\pi p_{\rm eff}/r^{2}$. The analysis treats the Horndeski geometry as a fixed background and focuses on the orbital dynamics and the corresponding leading-order gravitational-wave phenomenology.

The horizon structure depends directly on the effective Horndeski parameter $p_{\rm eff}$. For $p_{\rm eff}>0$, the spacetime possesses a single positive horizon, whereas for $-m^{2}/(8\pi)<p_{\rm eff}<0$ the metric admits two positive roots. The two roots merge at $p_{\rm eff}=-m^{2}/(8\pi)$, defining the extremal limit, while for smaller values no real horizon exists. The equatorial effective potential was used to determine the innermost stable circular orbit (ISCO) and the marginally bound orbit (MBO). Both $r_{\rm ISCO}$ and $r_{\rm MBO}$ increase monotonically with $8\pi p_{\rm eff}/m^{2}$. In particular, $r_{\rm ISCO}/m$ changes from $5.1665$ at $8\pi p_{\rm eff}/m^{2}=-0.5$ to $6.6945$ at $+0.5$, with the Schwarzschild value $r_{\rm ISCO}/m=6$ recovered at $p_{\rm eff}=0$.

Periodic bound orbits were classified using the Levin--Perez--Giz (LPG) taxonomy $(z,w,v)$. For the representative family considered here, $w=1$, $v=0$ for $z=1$, and $v=z-1$ for $z>1$, with $z=1,\ldots,6$. The resulting periodic trajectories contain exactly $z$ distinct radial leaves. At fixed $L_i$, the rational parameter
$q=\Delta\varphi_r/(2\pi)-1$
increases monotonically with the specific energy $E_i$ over the allowed interval $E_{\min}<E_i<E_{\max}$. In the nearly circular limit, $q$ approaches a finite lower value, with $q_{\min}\simeq0.6550,\,0.6374,\,0.6321$ for $8\pi p_{\rm eff}/m^{2}=-0.5,\,0,\,+0.5$, respectively. For the selected periodic family, the single-leaf orbit $(z,w,v)=(1,1,0)$ has $q=1$, while the cases with $z>1$ satisfy $q=2-1/z$. Nearby trajectories away from exact rational commensurability lose exact orbital closure, illustrating the special role played by rational orbits within the bound orbital phase space.

The gravitational-wave signals were estimated in the test-particle and leading-order quadrupole approximations for an EMRI configuration with central mass $m=4\times10^{6}\,M_{\odot}$, secondary mass $m'=100\,M_{\odot}$, and source distance $D_L=8000\,\mathrm{pc}$. The resulting time-domain polarizations and characteristic strain spectra exhibit the periodic structure inherited from the rational orbital motion. For the representative configurations considered here, the dominant spectral features lie in the mHz band, and part of the characteristic strain spectrum overlaps with or exceeds the nominal sky-averaged LISA sensitivity level. This comparison is intended only as an order-of-magnitude indication of the relevant frequency and strain scales; a quantitative assessment of detectability requires a full signal-to-noise-ratio calculation together with an accurate detector response and strong-field waveform model.

Overall, the effective Horndeski parameter $p_{\rm eff}$ produces systematic changes in the characteristic orbital radii, the rational-orbit structure, and the associated gravitational-wave spectra. These results provide a phenomenological characterization of rational-orbit dynamics in this class of Horndeski black-hole spacetimes and illustrate how modified-gravity effects may leave imprints on strong-field orbital motion and gravitational-wave signals. A more complete treatment will require perturbations of the Horndeski background together with radiation-reaction and self-force effects, which are beyond the scope of the present analysis.

\acknowledgments
This work was supported by the National Natural Science Foundation of China under Grant No. 12405154.

\providecommand{\href}[2]{#2}\begingroup\raggedright\endgroup

\bibliographystyle{JHEP}

\end{document}